\documentclass[floatfix,aps,prd,twocolumn]{revtex4-1}

\usepackage{amsmath}
\usepackage{amsfonts}
\usepackage{amssymb}
\usepackage{bm}
\usepackage{enumerate}
\usepackage{epsfig}
\usepackage{graphicx}
\usepackage{epstopdf}

\usepackage{pifont}
\usepackage{bbding}

\usepackage{mathptmx}   

\begin{document}

\title{Are Electron Scattering Data Consistent with a Small Proton Radius?}

\newcommand*{\WM}{Physics Department, College of William and Mary, Williamsburg, Virginia 23187}
\newcommand*{\WMindex}{1}


\author{Keith Griffioen} 

\author{Carl Carlson} 

\author{Sarah Maddox} 
\affiliation{\WM}
\date{\today}

\begin{abstract}
We determine the charge radius of the proton by analyzing the published low momentum transfer 
electron-proton scattering data from Mainz.  We note that polynomial expansions of the form 
factor converge for momentum transfers squared below $4m_\pi^2$, where $m_\pi$ is the pion mass.  
Expansions with enough terms to fit the data, but few enough not to overfit, yield proton 
radii smaller than the CODATA or Mainz values and in accord with the muonic atom 
results.  We also comment on analyses using a wider range of data, and overall  
obtain a proton radius $R_E=0.840(16$) fm.

\end{abstract}

\maketitle


\section{Introduction}

Much remains to be learned about the proton.  After a half-century of study, we still do not know 
what its size is, where its spin comes from, and how its mass is generated from light quarks and 
gluons. Particularly troubling is the matter of the proton's charge radius, $R_E$.  This was first 
measured to be approximately 0.8 fm by Hofstadter and collaborators in the 1950s via elastic electron 
scattering \cite{Chambers:1956zz}. The value of $R_E$ has been steadily refined over the years 
through electron scattering and hydrogen energy level measurements, recently reviewed in 
Refs.~\cite{Pohl:2013yb,Carlson:2015jba}. The {CODATA} group \cite{Mohr:2012tt,codata2014}, 
using available electron-based data through 2014, quotes a combined value of $R_E=0.8751\pm 0.0061$. 
The recent electron scattering experiment in Mainz~\cite{Bernauer:2010wm, Bernauer:2013tpr,Bernauer:2011zza}, 
which quotes a value of $R_E=0.879\pm 0.008$, is included in this {CODATA} value. 
For many years $R_E$ had remained relatively stable at about $0.88$ fm, until this 
value was called into question by Lamb shift measurements in muonic hydrogen, which 
yielded a value of $R_E=0.84087\pm 0.00039$ fm \cite{Pohl:2010zza, Antognini:1900ns}.  
This radius is 7 standard deviations away from the {CODATA} value.  The proton size puzzle leaves 
us with three options: the hydrogen Lamb shift and elastic electron scattering experiments have erred 
in extracting $R_E$, the muon Lamb shift measurement is precise, but inaccurate, or there is 
new physics that affects the muon differently than the electron, rendering the theory behind 
the muonic Lamb shift calculations incomplete \cite{TuckerSmith:2010ra,Batell:2011qq,
Barger:2011mt,Carlson:2012pc,Carlson:2013mya,Glazek:2014ria}.

In this paper, we explore whether the published, high-quality $ep$ elastic scattering data at 
low-$Q^2$ from Mainz could be consistent with the muonic Lamb shift 
determination of $R_E$. Extracting $R_E$ from elastic electron scattering is as 
simple, or as difficult, as measuring the slope of the electric form factor 
$G_E(Q^2)$ as a function of the squared four-momentum transfer $Q^2$ as $Q^2$ 
goes to zero. However, since the differential cross section diverges at small 
scattering angles (low $Q^2$), these measurements are very sensitive to beam 
alignment and angle determination. No $ep$ measurement extends to $Q^2=0$, although 
some get close.  Mainz currently holds the record, with measurements at $Q^2$ as 
low as $0.0038$ GeV$^2$.  This modern data set, with 1422 data points  
in the range from the lower limit to about $0.98$ GeV$^2$, is the best, 
most precise, and most extensive available.   Therefore, it is the Mainz data that 
we choose to explore.  

The charge radius is given by the second term in the expansion of the electric form factor,
\begin{align}
G_{E}(Q^2) = 1 - \frac{1}{6} R_E^2 Q^2 + c_2 Q^4 + \ldots	.
\end{align}
Using data at very low $Q^2$, one can hope the $Q^4$ term, the curvature term, is small so 
that the charge radius $R_E$ can be determined without having to model the shape of $G_{E}$ 
over a wider range of $Q^2$.  In the past, the size of the uncertainties on 
the data at very low $Q^2$ has meant that one could not extract an accurate charge 
radius without extending the data range to include not-so-low $Q^2$.  The $Q^4$ term 
then can become noticeable, depending on how far the data range is extended. 
There was an early example of Simon \textit{et al.}\ data~\cite{Simon:1980hu} 
where the fitted coefficient of the $Q^4$ term was small, albeit with large uncertainty, 
and the extracted charge radius was also small.  Using this example, some workers
({\it e.g.,}~\cite{Sick:2011lba}) 
advocated including data at still 
higher $Q^2$ to obtain a larger curvature.  This also led to a larger extracted proton 
radius.  Including higher $Q^2$ does not have to mean including data at all available $Q^2$, 
and we have the example of Ref.~\cite{Sick:2003gm} using
only data with $Q^2 < 0.62$ GeV$^2$ ($Q < 4$ fm$^{-1}$).   

Sick and Trautmann~\cite{Sick:2014sra}, in fact, suggest that data 
from $0.014$ to $0.056$ GeV$^2$ in $Q^2$ (or $Q$ from $0.6$ to $1.2$ fm$^{-1}$) 
is most crucial for finding the proton radius.   The consideration of what data range 
is sensitive to the proton size, given the foregoing discussion, must depend on the 
accuracy and precision of the data.  As the data improve, the 
range of $Q^2$ needed to obtain the proton radius will decrease. With the new 
Mainz data now available, we will explore the possibility that we can obtain a 
good proton radius result using only data with a rather low maximum $Q^2$.  

The Mainz data~\cite{Bernauer:2010wm,Bernauer:2013tpr,Bernauer:2011zza} data enjoy 
state-of-the-art radiative and Coulomb corrections.  One of the three spectrometers was 
used as a luminosity monitor to control systematic uncertainties, and the other two spectrometers 
measured separate kinematic points simultaneously. The data are dominated by 
point-to-point systematic uncertainties from background subtractions, drift-chamber 
inefficiencies, normalization factors, angle determinations, and the afore-mentioned 
corrections.  The slight leeway we have in fitting these data is to make small 
rescalings of the 34 normalization sets in the experiment and to enlarge the point-to-point 
error bars if the fluctuations of the data indicate that the quoted uncertainties are too small.  
We limit our form factor fits to the Mainz data set because
the systematic differences 
between separate experiments can introduce systematic effects in global fits, and  
the 1422 Mainz points already dominate the sample of world data below $Q^2=1$ GeV$^2$.

We shall advocate for fits using the 243 Mainz data points with $Q^2$ below $0.02$ GeV$^2$.  
In addition to the general principle that using only very low $Q^2$ will free us from model 
dependence incurred in extending the fits to higher $Q^2$, we are also using only data from a 
limited number of spectrometer settings. This substantially frees the data from any drift arising 
from the normalization adjustments that reconciled data from different spectrometer settings, 
and averts overfitting of inflections or statistical fluctuations in the data that can 
happen when using fit functions that contain many fit constants. 

In the following, Sec.~\ref{sec:form} presents the formalism pertinent 
to our discussion; Sec.~\ref{sec:lowq2} presents the extraction and discussion of the 
proton radius using low $Q^2$ data;  Sec.~\ref{sec:full} presents a fit based
on the full $Q^2$ range of the Mainz data; Sec.~\ref{sec:fr} highlights our 
final results; and Sec.~\ref{sec:conclusions} gives our conclusions.


\section{Formalism}

\label{sec:form}

When an electron of energy $E$ scatters from a proton at rest through an angle 
$\theta$ and exits with energy $E^\prime$, the 4-momentum transfer squared is,
\begin{equation}
Q^2 = -q^2 = 4 E E^\prime \sin^2\frac{\theta}{2}.
\end{equation}

In the Born approximation, the $ep$ elastic scattering cross section can be written in 
term of the electric and magnetic Sachs form factors, $G_E(Q^2)$ and $G_M(Q^2)$,
\begin{equation}
\frac{d\sigma}{d\Omega} = \left(\frac{d\sigma}{d\Omega}\right)_{\rm Mott}
	\frac{1}{(1+\tau)}
		\Big[  G_E^2(Q^2) +  
\frac{\tau}{\epsilon} \, G_M^2(Q^2)  \Big].
		\label{eq:rosenbluth}
\end{equation}
The Mott cross section is
\begin{equation}
\left(\frac{d\sigma}{d\Omega}\right)_{\rm Mott} =
	\frac{  4 \alpha^2 \cos^2\frac{\theta}{2}  }{ Q^4 }
		\frac{ {E^\prime}^3 }{ E },
\end{equation}
$\alpha$ is the fine structure constant, 
\begin{align}
\epsilon &= \left( 1 + 2(1+\tau)\tan^2\frac{\theta}{2}\right)^{-1},\,\,	{\rm and} \nonumber\\
\tau &= \frac{\nu^2}{Q^2} = \frac{Q^2}{4M^2 }		\,,
\end{align}
where $\nu$ is the energy transferred by the virtual photon and
$M$ is the proton mass.
Further,
\begin{equation}
E^\prime = \frac{E}{1 + ({2E}/{M})\sin^2\frac{\theta}{2}}		\,.
\end{equation}

The electric and magnetic form factors at $Q^2=0$ are normalized to correspond to the 
nucleon charge in units of $e$ and the nucleon magnetic moment in units of the proton 
magneton $\mu_N = e/(2M)$, such that
\begin{align}
G_{E}(0) = 1,\quad{\rm and}\quad   G_{M}(0) =\mu_p \approx 2.793.  
\end{align}

The dipole form factor 
\begin{equation}
G_D = \frac{1}{\left(  1+{Q^2}/{0.71\ \text{GeV}^2}  \right)^2},
\end{equation}
suitably normalized,
has been used for many years as a benchmark approximation for both
$G_E$ and $G_M$.  The Mainz group presents data on the cross section $\sigma$ as its ratio 
to the cross section calculated using the dipole form factors,  $\sigma_D$,
\begin{equation}
\frac{\sigma}{\sigma_{D}} = \frac{\epsilon G_E^2 + \tau G_M^2}
			{\epsilon G_D^2  + \tau \mu_p^2 G_D^2} 	\,.
	\label{eq:extract}
\end{equation}
From this,
\begin{align}
G_E(Q^2) = &G_D(Q^2) \left( \frac{\sigma}{\sigma_D} \right)^{1/2}	\nonumber\\
	&\times \left[
	1 + \tau\mu_p^2 \frac{ G_M^2/(\mu_p G_E)^2 - 1}{ \epsilon + \tau \mu_p^2  }
	\right]^{-1/2}	\,.
\label{eqn:10}
\end{align}

For $Q^2 < 0.02$ GeV$^2$, the quantity in square brackets above, for reasonable values of the 
$G_E/G_M$ ratio,  differs from unity by no more than 140 parts per million, and plays no 
significant role in the extraction of $G_E$.  For general $Q^2$, we most often will obtain 
$G_E$ using the $G_E/G_M$ ratio obtained from recoil polarization experiments, mostly at higher 
$Q^2$.  The data from JLab indicate that at least for $Q^2<8$ GeV$^2$, 
\begin{equation}
\mu_p \frac{G_E}{G_M} \approx 1-\frac{Q^2}{8\ \text{GeV}^2}	\,.
\label{eq:GEGM}
\end{equation}

Returning to low $Q^2$, one can use the radius expansion,
\begin{align}
\mu_p\frac{G_E}{G_M} &= 1 - \frac{1}{6}(R_E^2-R_M^2)Q^2 \label{eq:EM}	\,,
\end{align}
where $R_M$ is the magnetic radius and
\begin{align}
G_E = G_D \left( \frac{\sigma}{\sigma_{D}} \right)^{1/2}
	\left\{	1 +  \frac{\mu_p^2 Q^4}{12 M^2}
	\frac{ R_E^2 - R_M^2 }{ \epsilon + \tau \mu_p^2  }
	\right\}^{- {1}/{2}}  \,.
\end{align}
Specifically, Bernauer {\it et al.}~\cite{Bernauer:2010wm} obtain  $R_E=0.879\pm 0.008$ fm and $R_M=0.777\pm 0.017$ fm 
from their fits.  The latter is significantly smaller than most other fits, where typically $R_M$ is 
similar to $R_E$, but none-the-less, the extraction of $G_E$ from the data at low $Q^2$ 
is unaffected by the spread of suggested values for $R_M$ even at the high level of accuracy 
needed for the present investigation.


\section{Analysis for $Q^2<0.02$ GeV$^2$}

\label{sec:lowq2}

Looking only at data for $Q^2$ below $0.02$ GeV$^2$, there are a plethora of data points 
from the Mainz experiment. In this $Q^2$ range,  the term in brackets in Eq.~\ref{eqn:10} is, for all 
practical purposes, unity.  Mainz has 243 data points for $Q^2 < 0.02$ GeV$^2$. 
Limiting consideration to only spectrometer B to minimize cross 
calibration uncertainties, still leaves 209 data points.  Although the overall normalization 
may be uncertain by $1$--$2$\%, any relative systematic uncertainties that could lead to 
a false $Q^2$ dependence are thought to be small.

\begin{figure}[h]
\includegraphics[width= 8.3 cm]{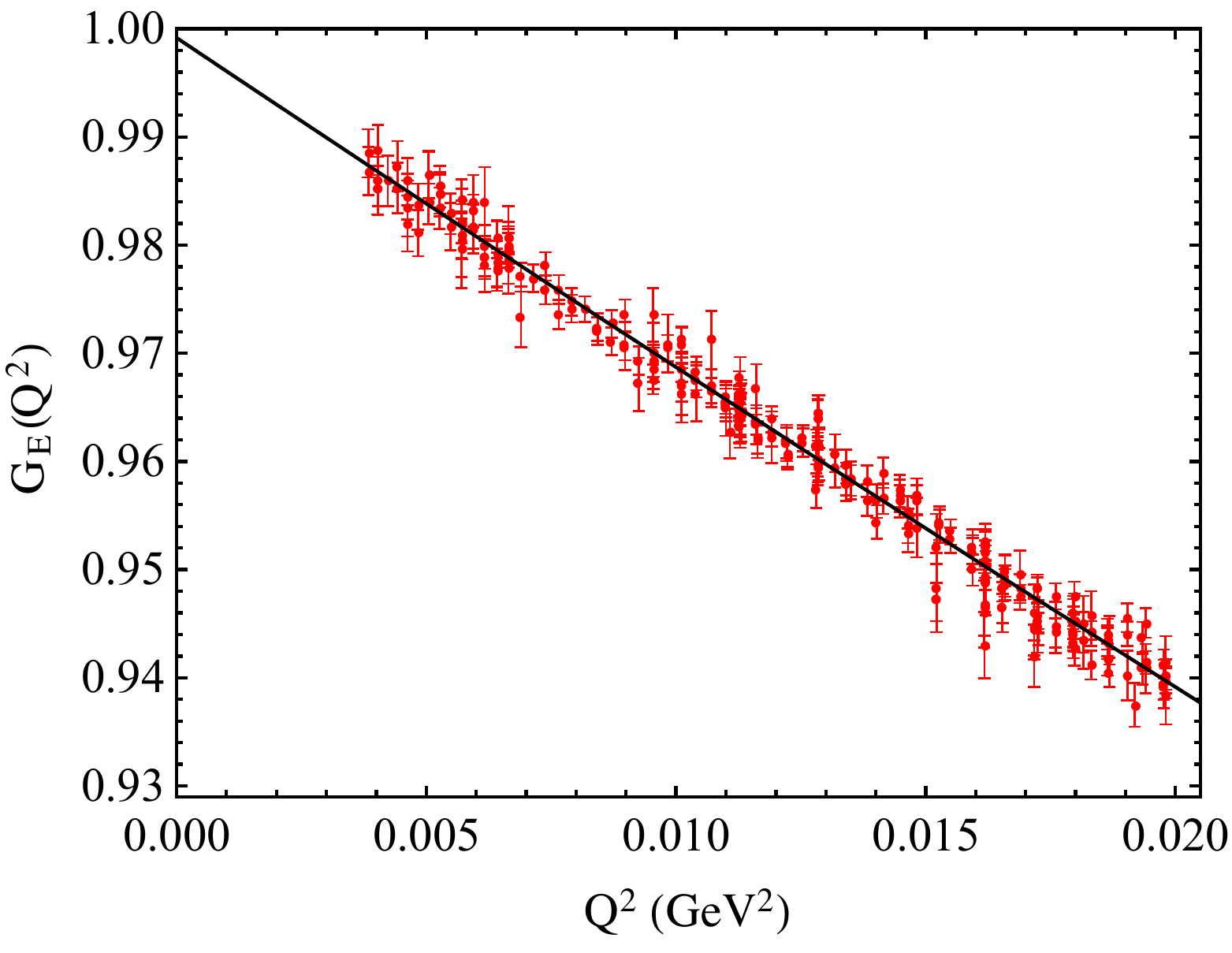}
\vskip -5mm
\caption{
(Color online) Linear plus quadratic fit to all Mainz (2010) data with $Q^2<0.02$ GeV$^2$.
}
\label{fig:radius0.02P2}
\end{figure}

We can fit the data using a linear plus quadratic in $Q^2$ form for $G_{E}$, 
\begin{equation}
G_{E}(Q^2) = c_1 (1 + c_2 Q^2 + c_3 Q^4) , 
\end{equation}
where $c_2 = - R_E^2/6$.  Using all 
available points below $0.02$ GeV$^2$ gives the result shown in Fig.~\ref{fig:radius0.02P2}.  
The $\chi^2$ per degree of freedom (dof) for the fit is 1.00, the normalization constant 
is $0.9992\pm 0.0003$,  $R_E = 0.850\pm 0.019$ fm, consistent with the Lamb 
shift results, and $c_3 = 4.5 \pm 5.6$ GeV$^{-4}$.  The central value of $c_3$ 
is positive as one might expect from a nonrelativistic expansion of the form factor, 
but is statistically consistent with zero.

Without prior expectations, one could use a rule of thumb for fitting, namely to discard 
terms in the fit that do not improve the $\chi^2/$dof.  A linear fit, $G_{E}(Q^2) = c_1(1+c_2 Q^2)$, 
leads to the same $\chi^2/$dof = $1.00$, and values $c_1 = 0.9986 \pm 0.0003$ 
and $R_E = 0.835 \pm 0.003$ fm (using the diagonal term in the error matrix).  From 
a statistical viewpoint, a radius as large as 
$0.88$ fm is unfavored.  However, the correlations are such that the more positive the 
curvature, the larger the extracted proton radius.

Fits like the ones just presented have been criticized because of expectations that the 
curvature could be larger than the apparent results from the quadratic fit and 
certainly not zero as in the linear fit.  To investigate the effects of curvature when 
fitting a low-$Q^2$ data set, we expand the electric form factor as
\begin{equation}
G_E(Q^2) = 1-\frac{1}{6} R_E^2Q^2 + \frac{b_2}{120}R_E^4Q^4 - \frac{b_3}{5040}R_E^6Q^6	,
\end{equation}
where the coefficients are suggested by nonrelativistic models, where $R_E$ is the rms 
proton radius, $R_E^2 = \langle r^2 \rangle$, and $b_2 =\langle r^4\rangle / \langle r^2\rangle ^2 $,
and  $b_3=\langle r^6\rangle / \langle r^2\rangle ^3$.  The coefficients can be calculated using 
exponential, Gaussian, and uniform model charge distributions $\rho(r)=\rho_0 e^{-r/a}$,
$\rho(r)=\rho_0 e^{-r^2/b^2}$, and $\rho(r)=\rho_0 \theta(c-r)$.  For these three cases, 
respectively, $b_2 = 5/2$, $5/3$, and $25/21$, and
$b_3 = 35/3$, $35/9$, and $125/81$.  The fits using $Q^2<0.02$ GeV$^2$ data yield 
$R_E= 0.859(3)$, $0.851(3)$, and $0.846(4)$ fm, respectively, each with a 
$\chi^2$/dof $= 1.00$.  One of these results is almost neutrally between the muonic 
and electronic radius values, while the Gaussian and uniform distributions even with 
the pre-chosen curvature term give results commensurate with the muonic Lamb shift value.

\begin{figure}[h]
\includegraphics[width= 8.2 cm]{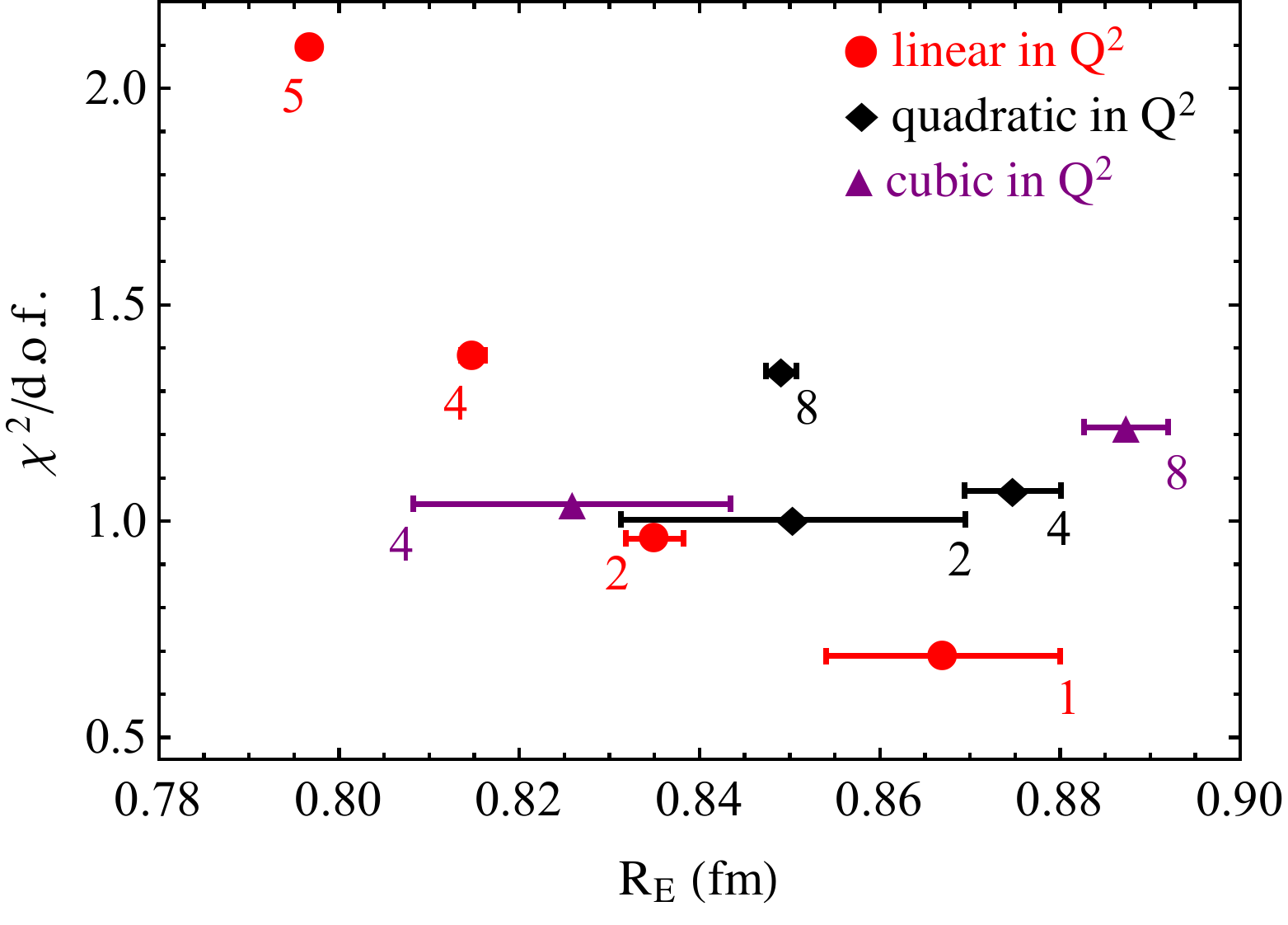}\\
\vskip -4 mm
\caption{(Color online)
$\chi^2$/dof vs.\ $R_E$ for polynomial fits to the data set, showing  fits with first, second, 
and third order polynomials in $Q^2$ using data with $0.01 < Q^2_{\max} < 0.08$ GeV$^2$.  
The small numbers near each data point gives $Q^2_{max}$ in multiples of 
$0.01$ GeV$^2$.  
A good fit should have a good $\chi^2$ and a sufficient but not oversufficient number of parameters,
as further discussed in the text.  The quadratic fit ``2'' and the cubic fit ``4'' satisfy the criteria.
}
\label{fig:polylow}
\end{figure}

The fit just discussed is one example.  One may inquire what results follow with different 
$Q^2_{max}$ (always with $Q^2_{max} < 4 m_\pi^2$) and different orders of polynomial.  
There will be two criteria for an acceptable fit.  One is that the $\chi^2$ is 
low enough, on the order of 1 per degree of freedom.  The other is that the 
highest order term in the polynomial is not well determined, as judged by the 
uncertainty limit on its coefficient, with  the previous terms well determined.  
This will imply that the fit has omitted no important term, and is good. 
Fig.~\ref{fig:polylow}  shows $\chi^2$/dof and $R_E$ from a number of 
linear, quadratic, and cubic (in $Q^2$) fits, with the small numbers indicating $Q^2_{max}$ 
for each example in multiples of $0.01$ GeV$^2$.

The linear fits cannot sensibly satisfy the second criterion and still give a radius, but they are 
included to show what happens when $Q^2_{max}$ increases and the number of terms in the fit does not.  
The diagonal elements of the error matrix are astonishingly small, but the fits are poor, as 
judged by $\chi^2$.  Three quadratic fits are shown, with $Q^2_{max}$ indicated on the Figure.  
The lowest one is our prime example, which satisfies all criteria.  It has an acceptable $\chi^2$ 
and a small contribution from the actual $Q^4$ term in the polynomial.  The next still has an 
acceptable $\chi^2$, but the coefficient of the $Q^4$ term is stringently determined 
(circa $10\%$ uncertainty), leading to worries that a cubic expansion is not flexible enough for this 
$Q^2_{max}$, and so should not be trusted.
The last point in this series sees some rise in $\chi^2$, and with tightly determined 
coefficients again indicating underfitting at this $Q^2_{max}$.  For the two cubic fits shown, 
the one with lower $Q^2_{max}$ reasonably satisfies the criteria, while the upper one has 
its $Q^6$ coefficient tightly determined, again indicating underfitting.  Quartic fits did not 
give suitable results with $Q^2_{max}$ limited by the $\pi\pi$ threshold.  Some $Q^2_{max}$  
values, although happening to give an $R_E$ that matches the muonic hydrogen results, leads to fits with several 
poorly determined coefficients, and for other $Q^2_{max}$ the sign pattern of the coefficients does
not match the alternation expected for smooth charge distributions and the Fourier transform formula.  
The overall conclusion from further examinations of fits to the the low $Q^2$ range of 
the data, is that when the criteria for good fits are satisfied, the proton radius accords with the 
muonic hydrogen value.


\section{The Full $Q^2$ Range}

\label{sec:full}

We have so far concentrated on using low-$Q^2$ data to obtain the proton radius, but there is 
interest in considering the full data set.  
There are several topics to discuss.  Can a smooth function, with relatively few 
parameters give an acceptable fit to the data, and  what is the radius that follows from such a fit?  
What is the value and effect of fitting with more parameters?  
With more 
parameters one can fit systematic deviations or statistical fluctuations from what is 
really smooth data, which can mar the overall fit and skew the extrapolation to the proton radius.  
Also, with more parameters, there is a tendency for extensions 
outside the fit region to rapidly deviate from a 
properly smooth continuation of the data.  

Another item to consider is that
the polarization method for obtaining the form factors has shown that $G_E$ falls relative 
to $G_M$ with increasing momentum transfer, approximately as 
$\mu_p G_E/G_M = 1 - Q^2/(8$ GeV$^2)$~\cite{Jones:1999rz,Gayou:2001qd,Puckett:2010ac,Punjabi:2005wq,Puckett:2011xg}.  
We used this earlier when obtaining $G_E$ from the cross section data. For low $Q^2$ data the 
difference between $G_E$ obtained using the polarization results and using scaling, 
$\mu_p G_E = G_M$, is minor.  However, the full range of the Mainz data gives unexpected support for 
the polarization result. This is surprising, considering that it is a Rosenbluth experiment without
hard two-photon corrections, while all earlier Rosenbluth results gave scaling ($\mu_pG_E/G_M\approx 1$).  
Also in the absence of two-photon corrections, a reduced cross section at fixed $Q^2$ 
should be linear in $\epsilon$ (see Eq.~(\ref{eq:rosenbluth})).  Two-photon corrections change the slope 
in $\epsilon$, and may also give some $\epsilon^2$ and higher dependence, which will be sought in the data 
in the ensuing subsections.


\subsection{Full Range Fit}


For our analysis of the full data set, we have chosen a continued fraction (CF) form
\begin{equation}
f(Q^2)=\frac{c_1}{1+\frac{c_2 Q^2}{1+\frac{c_3 Q^2}{1+\frac{c_4 Q^2}{1+...}}}},
\end{equation}
for $G_E(Q^2)$,
in which $c_2 = (R_E /\hbar c)^2/6$.
A truncated continued fraction is a ratio of
polynomials, and it resembles Pad\'e approximates.  The continued fraction could be dangerous, because
it can have singularities in the spacelike region whenever one of the constants $c_i$ is negative.
However, if singularities do not occur within the fit range,  the continued fraction is acceptable,
and it allows a wide range of shapes covering several orders of magnitude with relatively few
parameters. On the other hand, fit functions with few parameters are unable to capture small inflections
in the data. 
Since there is no theoretical restriction that forbids inflections, many people believe
that they should exist.  We have looked for inflections in the data, and we find no 
persuasive argument to include them
in our parameterization. The low-$Q^2$ expansion of the continued fraction is
\begin{eqnarray}
f_{\rm lowQ}(Q^2)&=c_1[1-c_2 Q^2 + c_2(c_2+c_3)Q^4 \nonumber \\
& + c_2((c_2+c_3)^2 - c_3 c_4) Q^6 +...].
\end{eqnarray}

\begin{figure}[h]
\vskip -12 mm
\centerline{\includegraphics[width= 9.6 cm]{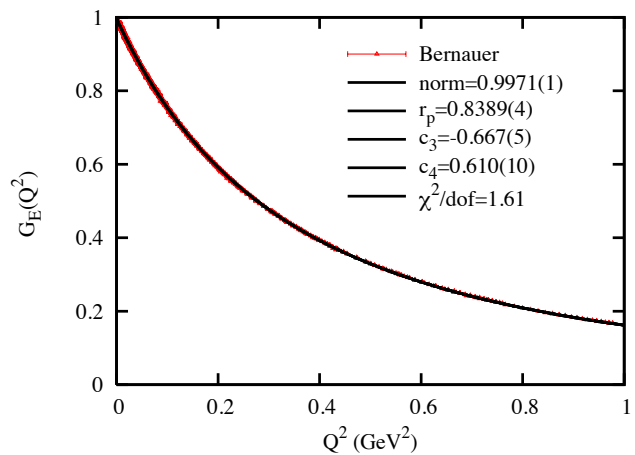}}
\vskip -8mm
\caption{(Color online)
Four-parameter continued fraction fit to the full Mainz data set using $\mu_p G_E/G_M=1-Q^2/(8{\rm\ GeV}^2)$.
}
\label{fig:radius1.0CF4}
\end{figure}

We extracted $G_E$ vs.\ $Q^2$ using Eq.~(\ref{eqn:10}) and fit all data to a 4-parameter continued fraction form.
Adding a fifth parameter did not improve the fit, so we limited ourselves to 4 parameters.  The
$\chi^2/$dof is 1.6, which is somewhat high, and we shall have more to say about this later.  However, the data are well-fit on average in all regions of $Q^2$.  From this fit we obtain $R_E=0.8389\pm 0.0004$.
The uncertainty on $R_E$ is small because a constraining 
fit form introduces information into the problem, in this case a belief in smoothness, which is in turn
reflected in the small diagonal uncertainties.  
Said another way, if the form factor can be faithfully described with a continued fraction with only a few terms,
then $R_E$ is tightly constrained.  But does the limited freedom of the continued fraction fit inappropriately force
a small value of $R_E$?
We shall also  discuss choosing other forms that may drive $R_E$ one way or the other, especially if we
allow undulation in an otherwise smoothly falling form factor.  We note that our value for 
$R_E$ is consistent with the analyses of the  Mainz data by Lorenz {\it et al.}\ using theoretically 
motivated analytic forms for $G_E$ \cite{Lorenz:2012tm,Lorenz:2014vha}.


\subsection{$G_E/G_M$}

\begin{figure}[h]
\vskip-11mm
\centerline{\includegraphics[width= 10.0 cm]{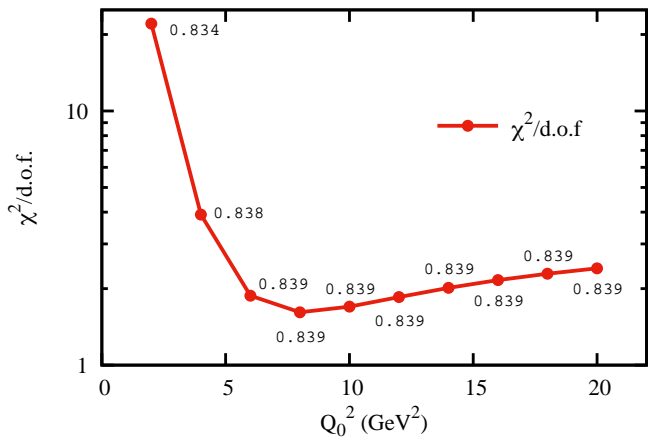}}
\vskip-11mm
\caption{(Color online)
Minimum $\chi^2/$dof\ values for continued fraction fits of $G_E$ extracted using $\mu_p G_E/G_M=1-Q^2/Q^2_0$
with different values of $Q^2_0$, when using the Mainz data.  
Each point shows the value of $R_E$ extracted for a given $Q_0^2$.
The quantity $R_E$ remains stable over the wide range $4<Q_0^2<20$ GeV$^2$, indicating that $R_E$ is
not sensitive to the {\it Ansatz} for $G_E/G_M$.
}
\label{fig:chisq_GEGM}
\end{figure}

We have extracted $G_E$ assuming that $\mu_p G_E/G_M=1-Q^2/Q_0^2$ for $Q^2 = 8$ GeV$^2$, and 
now wish to investigate consequences of different choices for $G_E/G_M$,  a least to the extent of 
considering other choices for $Q_0^2$. 
Fig.~\ref{fig:chisq_GEGM} shows the resulting $\chi^2$/dof  for various values of $Q^2_0$ upon 
fitting the 1422 data points with a 4-parameter CF function. The full data set favors a 
value of $Q^2_0\approx 8$ GeV$^2$.  This is bounded sharply on the low side and weakly on 
the high side.  The numbers shown beside each point are the values of the extracted
radius $R_E$ which are  only slightly influenced by the $G_E/G_M$ ratio used to extract 
$G_E$ from Eq.~(\ref{eq:extract}).    Moreover, the radius $R_E=0.84$ fm is stably reproduced for
$4  < Q^2_0 < 20$ GeV$^2$.  In particular, for steeper $\mu_p G_E/G_M$ slopes, the extracted 
$R_E$ actually decreases slightly.  This is a result rather different from Bernauer 
{\it et al.}, who obtain a larger radius and $\mu_p G_E/G_M=1-Q^2/(1.4$ GeV$^2)$ at very low $Q^2$.

\begin{figure}[h]
\vskip -2mm
\includegraphics[width= 8.6 cm]{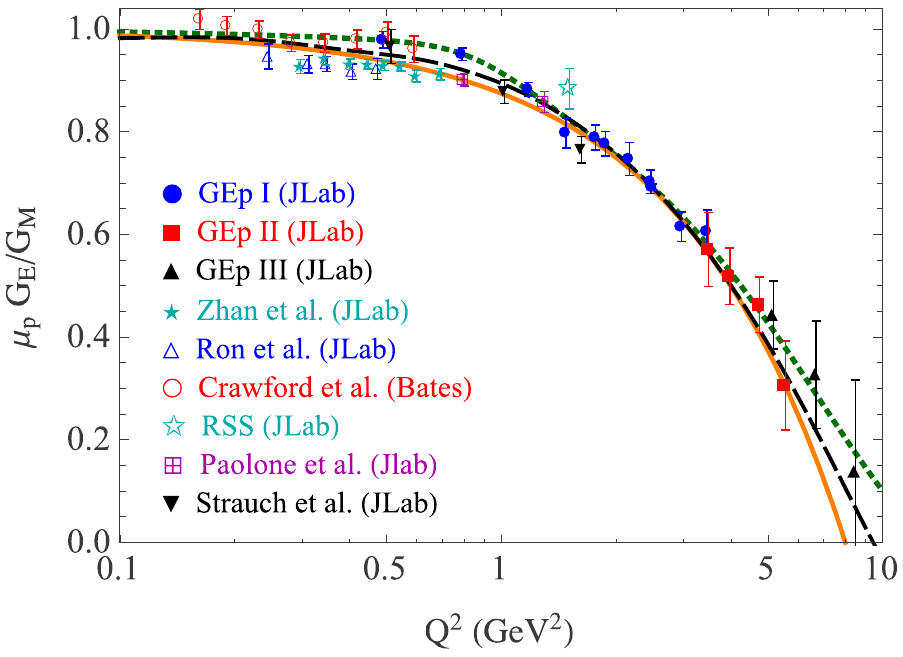}
\vskip -2mm
%
\caption{(Color online) 
Fits to world polarization transfer data.  The data are from GEp I~\cite{Jones:1999rz,Punjabi:2005wq}, 
GEp II~\cite{Gayou:2001qd,Puckett:2010ac}, GEp III~\cite{Puckett:2011xg}, Zhan 
\textit{et al.}~\cite{Zhan:2011ji}, Ron \textit{et al.}~\cite{Ron:2011rd}, 
Crawford \textit{et al.}~\cite{Crawford:2006rz},  RSS~\cite{Jones:2006kf}, Paolone \textit{et al.}~\cite{Paolone:2010qc}, and Strauch \textit{et al.}~\cite{Strauch:2002wu}.  
Some data with larger uncertainty limits have been omitted.  The solid orange line uses 
$\mu_p G_E/G_M=1-Q^2/Q_0^2$ with $Q_0^2 = 8.02$ GeV$^2$, the black dashed line 
is the Bernauer \textit{et al.}~\cite{Bernauer:2010wm, Bernauer:2013tpr,Bernauer:2011zza} 
fit up to about $0.3$ GeV$^2$ and a hybrid produced by other Mainz workers beyond 
that~\cite{Vanderhaeghen:2010nd}, and the green dotted line is a fit from Punjabi 
\textit{et al.}~\cite{Punjabi:2015bba}.
}

\label{fig:GEoverGM}
\end{figure}

Fig.~\ref{fig:GEoverGM} shows the world's polarization transfer data
\cite{Ron:2011rd,Ron:2011zz,Puckett:2011xg,Zhan:2011ji,Puckett:2010ac,Ron:2009zza,
Ron:2007vr,MacLachlan:2006vw,Punjabi:2005wq,Gayou:2001qd,Jones:1999rz}
for $\mu_p G_E/G_M$ on the proton (over a somewhat wider range than we have considered in the bulk of this paper).  It also shows the fit we have used, and  two other fits.   Incidentally, fitting the form
$f(Q^2)=1-Q^2/Q_0^2$ just to the data gives $Q_0^2=8.02\pm 0.05$ with $\chi^2/$dof =  2.3.   
Although the recoil polarization method is the best way we know to
determine the electric to magnetic form factor ratio, being relatively free of 
two-photon effects, the data points below $Q^2 \approx 0.8$ GeV$^2$ disagree with each 
other more than their quoted uncertainties would allow.   The variate $u_i = [y_i-f(x_i)]/\sigma_i$ 
has a small mean of 0.02 and a large standard deviation of 4.0.  From this we 
conclude that a linear fit acceptably represents the average of the data points,
despite their underestimated uncertainties. 

That the Mainz data also prefer $Q^2_0=8$ GeV$^2$, consistent with the recoil polarization data, 
leads to a conundrum.  
Earlier Rosenbluth results gave scaling.
The drop in $G_E/G_M$ with increasing $Q^2$ was a great surprise when 
announced in 1999 and published in 2000~\cite{Jones:1999rz}.  Why the Mainz data, without 
full hard two-photon corrections, agrees with the polarization results is a mystery.

\subsection{Epsilon Dependence}


\begin{figure}[h]
\vskip -9 mm
\includegraphics[width = 9.1 cm, angle=0]{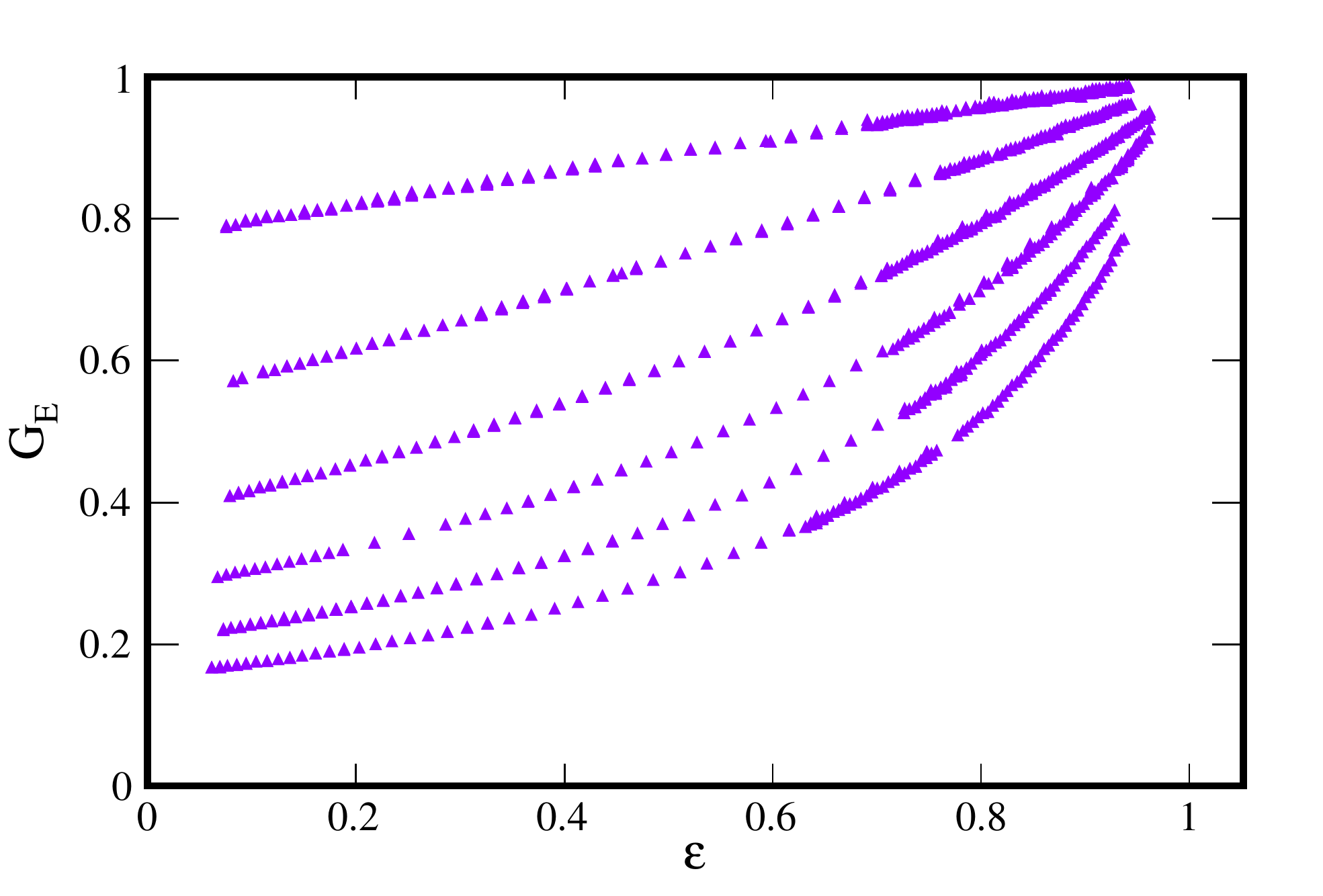}
\vskip -5 mm
\caption{(Color online) $G_E$ vs. $\epsilon$.  The six sets, from top to bottom, 
correspond to the six beam energies of the experiment,
180, 315, 450, 585, 720 and 855 MeV.
}
\label{fig:epsilon}
\end{figure}

Fig.~\ref{fig:epsilon} shows $G_E$ versus $\epsilon$, for the Mainz data,
for varying $Q^2$.  There are 6 sets of points
corresponding to the different beam energies of the experiment.  Since $G_E$ is a function of $Q^2$ (and the $G_E$ obtained from data will reflect this if all corrections are made), a horizontal line on this plot intersects the values of $\epsilon$ represented by points at fixed
$Q^2$.  Likewise, a vertical line on the plot shows different values of $Q^2$ at constant $\epsilon$.  The large
range in $\epsilon$ covered allows us to determine $G_E/G_M$, as discussed earlier.  It is also possible that some of the
apparent $\epsilon$ dependence can be attributed to mismatches in the normalization of the data from different beam energies.

 
\begin{figure}[h]
\vskip -3 mm
\includegraphics[width= 8.6 cm, angle=0]{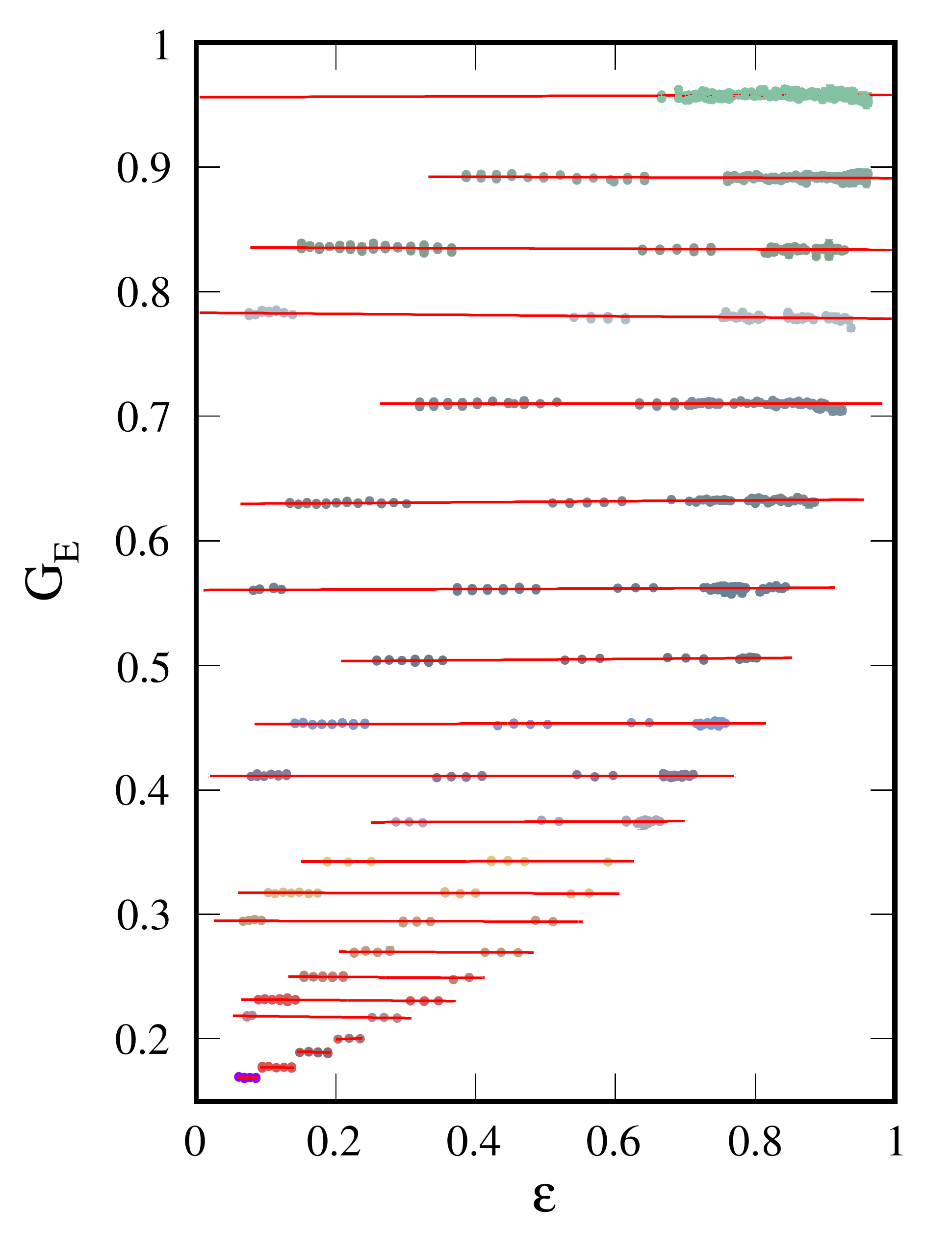}
\vskip -7 mm
\caption{(Color online)
Dependence of the extracted $G_E$ values on $\epsilon$ for 20 bins in $Q^2$.  The data show little or no dependence on $\epsilon$ within statistics.  (If desired, the actual $Q^2$ values may be inferred from the next Figure.)
}
\label{fig:ResidualEpsilon}
\end{figure}

Fig.~\ref{fig:ResidualEpsilon} shows the data set plotted versus $\epsilon$ for 20 bins in $Q^2$.  To ensure a common $Q^2$ for each horizontal line in this plot, data within a given $Q^2$ bin
were evolved to a central $Q^2$ using the fit $f(Q^2)$:
\begin{equation}
G_E(Q^2_{\rm average}) = G_E(Q^2_{\rm\ measured})\frac{f(Q^2_{\rm\ average})}{f(Q^2_{\rm measured})}.
\end{equation}

\begin{figure}[h]
\vskip -3 mm
\includegraphics[width=9.2cm, angle=0]{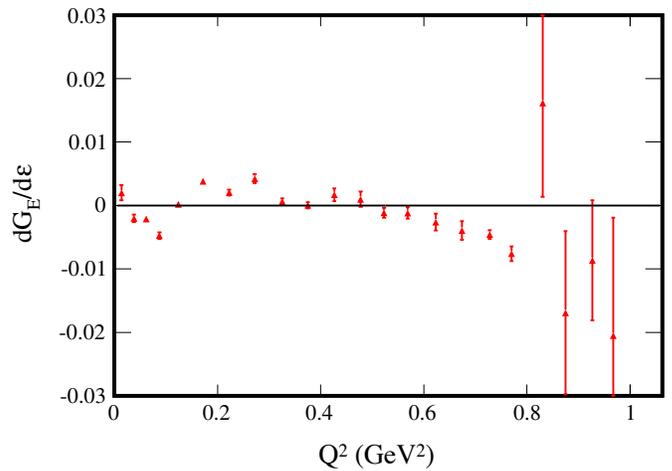}
\vskip -8 mm
\caption{(Color online)
$\epsilon$ slopes for 14 bins in $Q^2$.
}
\vskip -5 mm
\label{fig:SarahEps}
\end{figure}

The lines in Fig.~\ref{fig:ResidualEpsilon} look very flat, and the actual slopes of the data in Fig.~\ref{fig:ResidualEpsilon} are shown in Fig.~\ref{fig:SarahEps}.  On average, these slopes
are zero, but there are small variations.  At low $Q^2$, $G_M$ barely contributes to the cross section, so any 
$\epsilon$-slope must be an indication of a normalization mismatch in data from different beam energies.  The slopes are never more than a fraction of a percent, which can easily be accounted for by systematic variations in the data.


\subsection{Two Photon Contributions}


A possible cause of real $\epsilon$-dependence stems from two-photon exchange effects.  Although the Mainz data set 
includes Coulomb corrections following McKinley and Feshbach~\cite{McKinley:1948zz}, 
which are for the limit of very heavy pointlike protons, there are further 
hard two-photon effects occasioned by the hadronic structure of the 
proton~\cite{Blunden:2003sp,Chen:2004tw,Afanasev:2005mp}.  

Recent data on the cross section ratio of positron to electron elastic scattering from the proton 
verifies the idea that the Rosenbluth extraction of the $G_E/G_M$ ratio receives 
significant corrections from two-photon exchange~\cite{Rachek:2014fam,Adikaram:2014ykv}.

A potential further consequence of two photon exchange is that in addition to changing the $\epsilon$ 
slope in the reduced cross section $G_M(Q^2) + (\epsilon/\tau) G_E^2(Q^2)$,  there could also be 
terms quadratic or higher in $\epsilon$~\cite{Abidin:2007gn}.  
However, the data show that any $\epsilon^2$ terms are small.  
We conclude that two-photon corrections, although they have been demonstrated to exist \cite{Gorchtein:2014hla, Tomalak:2015aoa},
do not induce strong curvature in the Rosenbluth plot or bias the data in such a 
way as to change the radius $R_E$ if one does not have data over the full range of $\epsilon$.


\subsection{Fitting with Polynomials}       \label{sec:poly}

\begin{figure}[h]
\includegraphics[width= 8.4 cm]{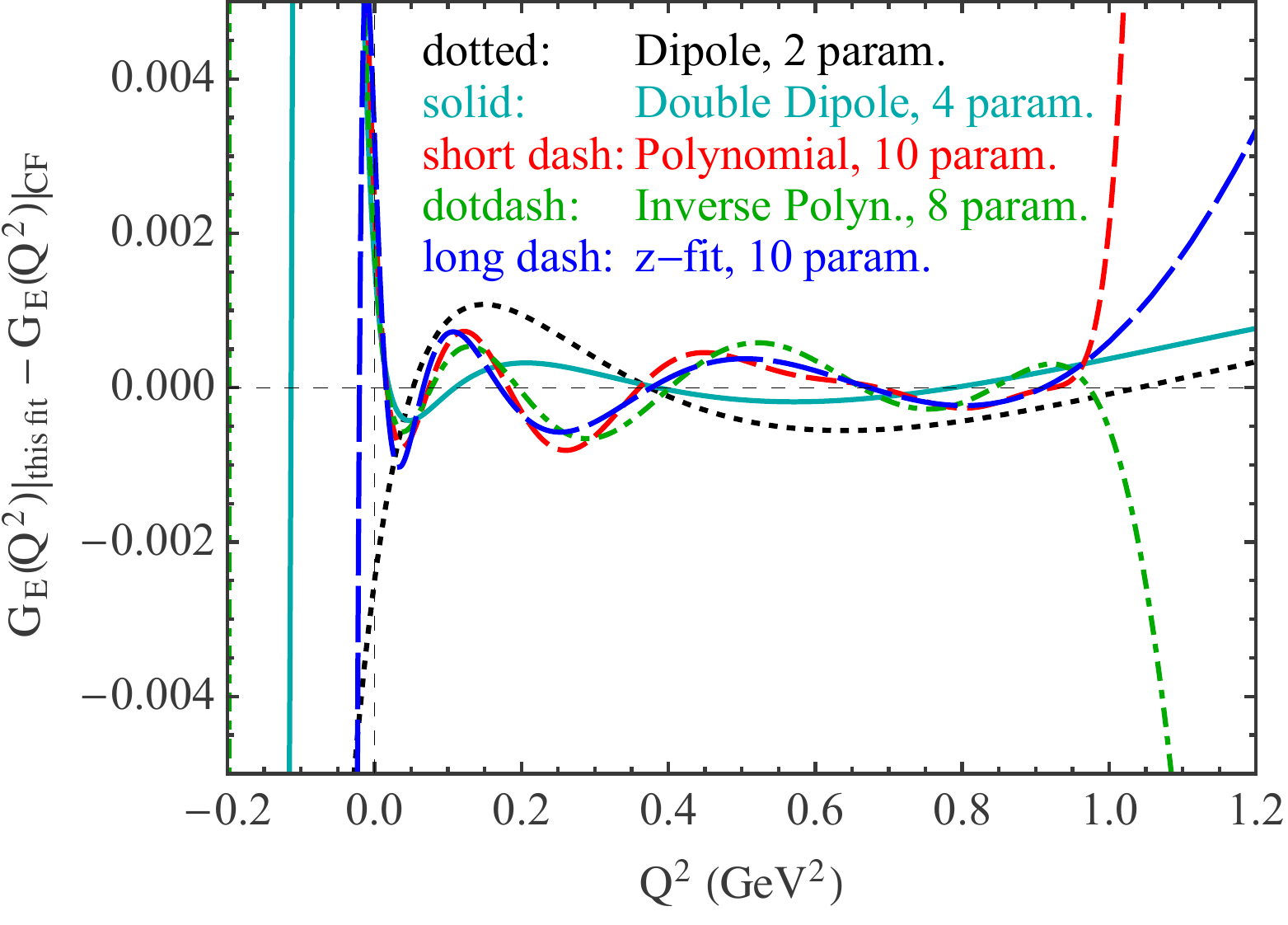}
\caption{(Color online)
Differences between various fit forms and the standard continued fraction fit.
All the power-series fit forms show undulations when enough parameters
are included.  
The $\chi^2$/dof drops accordingly to about 1.37, but at the expense of
pathological behavior at the origin and above $Q^2=1$ GeV$^2$.
}
\label{fig:FitCompare}
\end{figure}

The continued-fraction fit has only a few parameters and
may not accurately describe inflections, if there are physical inflections, of the 
measured form factor.  In this subsection, we experiment with other fits to the full 
range of the Mainz data.  We consider five different generic types of fit functions, 
in addition to the CF fit already presented, and will show some comparison of the different fits in 
Fig.~\ref{fig:FitCompare}.  See also 
\cite{Horbatsch:2015qda,Lee:2015jqa, Arrington:2015ria, Lorenz:2014yda,Graczyk:2014lba, Higinbotham:2015rja}.

First,  for reference, we fit the whole data set to a dipole form $a_0(1+Q^2/a_1)^{-2}$.  Although the $\chi^2$/dof
is larger than acceptable at $2.28$, the fit visually is remarkably good, and $R_E = 0.8299\pm 0.0002$ fm. 
The dipole is famous for giving a small radius when fit to a long data set, although this value is intriguingly close to the muonic hydrogen Lamb shift value of 0.841 fm.  

Second, we have followed the lead of Bernauer
{\it et al.}~\cite{Bernauer:2010wm} and fit to a double dipole  
$f(Q^2)=a_0[a_1(1+Q^2/a_2)^{-2} + (1-a_1)(1+Q^2/a_3)^{-2}]$.  
The fit has $\chi^2$/dof=1.6---the best apparently we can achieve with a 
smoothly and monotonically falling fit function---and $R_E=0.859\pm 0.001$ fm.

Third, we consider polynomial fits.  We do not advocate using polynomial fits beyond the 
spacelike reflection of the $\pi\pi$ threshold, since convergence of the fit is not 
assured beyond this point.  However, they have been used elsewhere, and we would 
like to comment on the results.  Polynomial fits with sufficient terms offer flexibility 
to fit inflections in the data, but they inevitably diverge outside any fit region, 
and accuracy at the end points is often poor for global fits.  

Fourth, we consider inverse polynomials $a_0/(1+\Sigma_{i=1}^N a_i Q^{2i})$.   

Fifth, we consider power series expansions in
 $z(Q^2)$, $f(z)= a_0(1 + \Sigma_{i=1}^N a_i z^{i})$, as advanced in this context by 
Hill and Paz \cite{Hill:2010yb}, where
\begin{equation}
z(Q^2) = \frac{\sqrt{4m_\pi^2 + Q^2}-2m_\pi}{\sqrt{4m_\pi^2 + Q^2}+2m_\pi}.
\end{equation}
The mapping to $z$ is motivated because a polynomial expansion of the form factor in $z$ 
converges for all spacelike $Q^2$, as long as the cuts or poles in the form factor are 
at timelike $q^2$ with $q^2 \ge 4 m_\pi^2$.

Fig.~\ref{fig:FitCompare} shows a visual comparison of five fits: the dipole fit, the double dipole fit, and representatives of the other three fit types.  The curves correspond to the differences between each model tested and the
4-parameter continued fraction (CF) fit described earlier.  
All polynomial fits show multiple oscillations around the CF value.
Moreover, the curves are clearly pathological near $Q^2=0$ and $Q^2=1$ GeV$^2$, that is to say, just outside the region where the fitted data has support.  With sufficient parameters,
the polynomial, inverse polynomial, and $z$-fits all start to reproduce inflections in the data, and they track each other roughly.  The large rise at $Q^2=0$ is the reason
these fits give a larger radius than the CF fit.  Although in absolute terms, the fits differ from each other by less than the point-to-point uncertainties on the
data, and absolutely less than 0.001, the precise behavior of the fit function at the origin significantly influences the extracted value of $R_E$.

\begin{figure}[h]
\vskip -8 mm
\includegraphics[width= 9.2 cm]{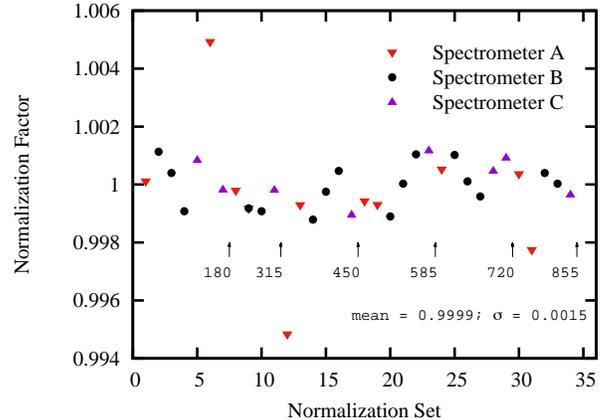}
\vskip -8 mm
\caption{(Color online)
Renormalization constants for the 34 Mainz normalization sets.  
Red inverted triangles, black dots, and purple upright triangles correspond to Spectrometers A, B, and C, 
respectively.  The numbers are the 6 beam energies in MeV, and the points to the left of the corresponding
arrow are the sets at that energy. 
The average and standard deviation of these normalization constants are 1 and 0.015\%, respectively.
}
\label{fig:renorm}
\end{figure}


\vskip -14mm
\subsection{Systematic Deviations Between Fit and Data}


\begin{figure}[h]
\vskip -2 mm
\includegraphics[width=9.2cm, angle=0]{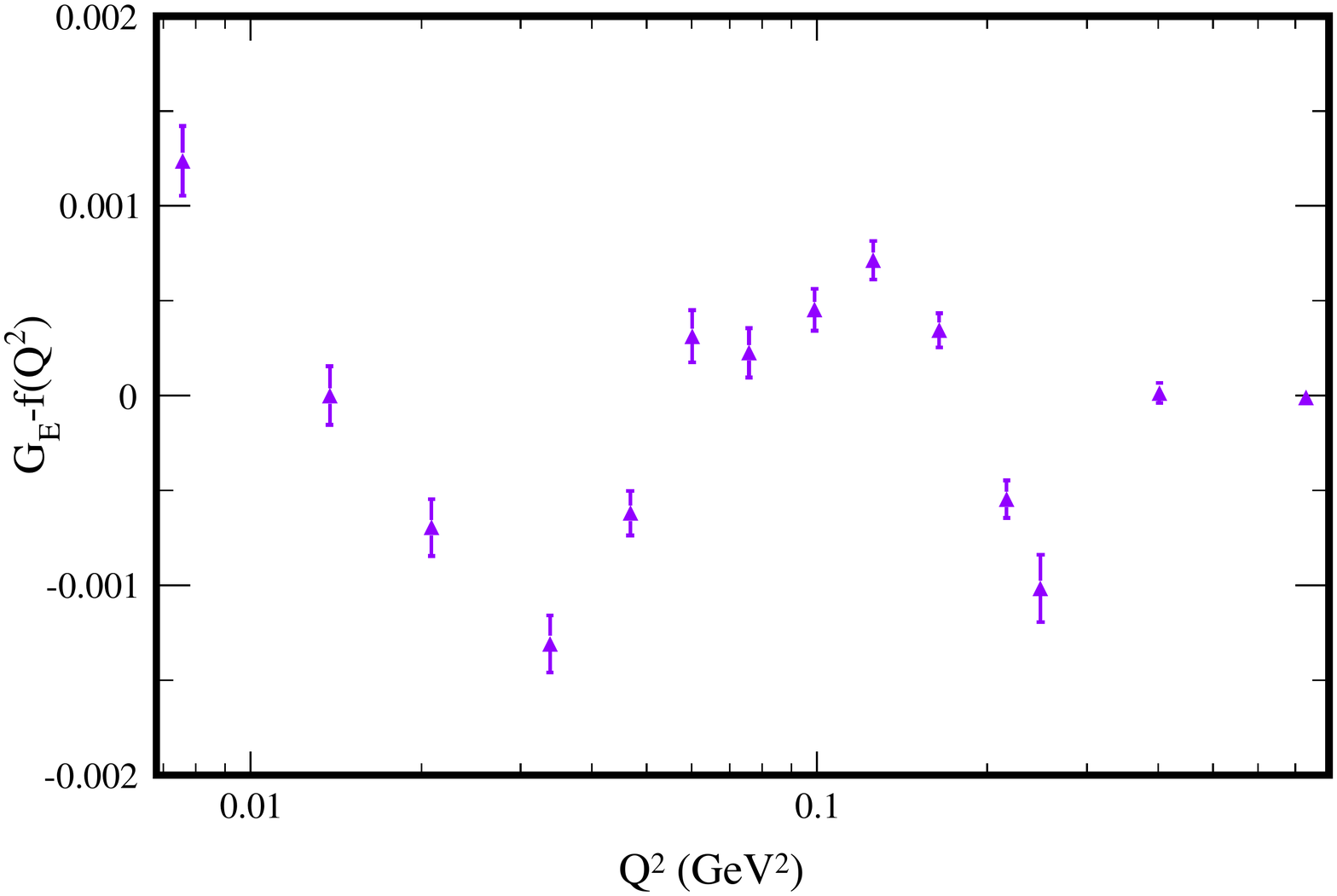}
\vskip -7 mm
\includegraphics[width=9.2cm, angle=0]{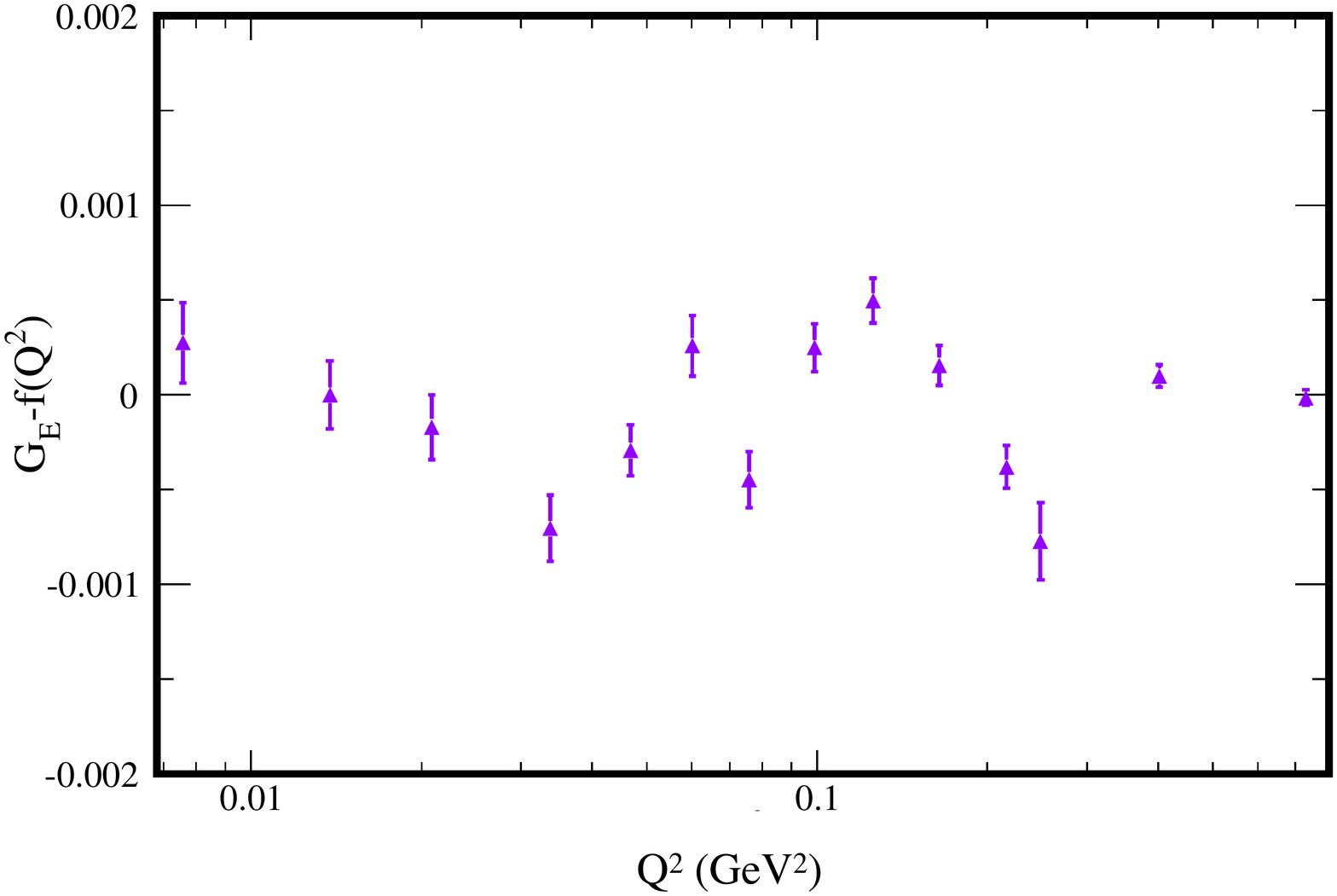}
\vskip -10 mm
\caption{(Color online)
Average differences $(G_E)_i - f(Q^2_i)$ for 14 $Q^2$ bins.
Upper: before renormalizing the 34 data sets; lower: after
renormalizing. 
}
\vskip -7mm
\label{fig:AveDiffBefore}
\end{figure}

The typical quoted uncertainty on each point is a few tenths of a percent.  Since these data represent 34 separately normalized
data sets taken with three spectrometers, it is not unreasonable to suppose that some of the apparent undulation 
could be modified or removed by relative renormalization. Since absolute normalizations in each
spectrometer are not known
to better than a percent, there is some freedom to do this on the level of at least a few tenths of a percent.

These sorts of relative renormalizations were made by Bernauer \textit{et al.}, with slightly 
different renormalizations for the different fit functions they used.  We did a similar process 
using the continued fraction fit.
For each normalization set we formed the uncertainty-weighted average of the ratio $G_E(Q^2_i)/f(Q^2_i)$
for all points in each subset.  The data were then divided by the inverse of this ratio.
These factors are shown in Fig.~\ref{fig:renorm} for the various data sets.  Arrows indicate 
the beam energy of the points to the left of the arrow.  The
overall renormalization is unity, with a point-to-point variation of about 0.15\%.  
This indicates that the original normalizations
were done well, although they could be be modified a bit when using a different fit function. 

Attempting make it easier to see any systematic effects within this thicket of renormalization ratios, we  
combined points from different spectrometer settings within the full data sample into 14 bins of $Q^2$ 
with roughly 100 points in each bin.  Fig.~\ref{fig:AveDiffBefore} shows the results of this exercise 
before the renormalizations were done (top) and after (bottom).   
Each  plot shows the average differences, $G_E(Q^2)-f(Q^2)$.  There are what appear 
to be statistically significant variations in $G_E(Q^2)-f(Q^2)$ in the upper plot.  In fact, focusing on 
the low-$Q^2$ behavior, the data show a trend favoring a larger slope, and a bigger radius $R_E$, than 
our fits suggest.  However, it is worth noting that these variations are on the order  of 0.1\%, which 
is commensurate with the point-to-point uncertainties.  In the lower plot, the systematic variations seen in the 
upper plot are reduced by half, but renormalization cannot account for the remaining fluctuations which are
on the order of 0.001.

\begin{figure}[h]
\includegraphics[width=9.2cm, angle=0]{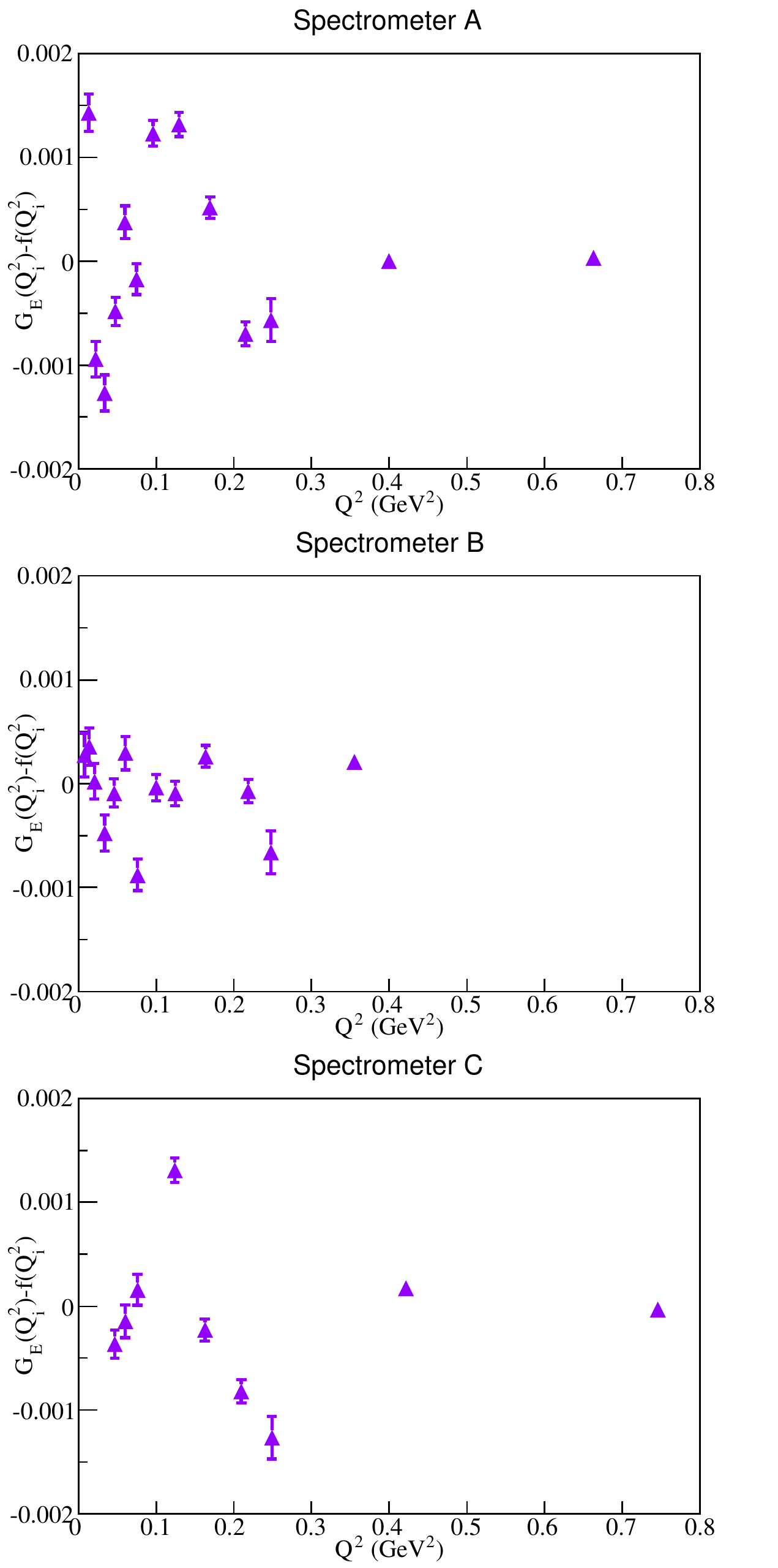}
\caption{(Color online)
Differences, $G_E(Q^2_i) - f(Q^2_i)$, averaged over 14 $Q^2$ ranges
for Spectrometers A, B, and C individually, after renormalizations of the 34 subsets.
}
\label{fig:DeltaABC}
\end{figure}

Fig.~\ref{fig:DeltaABC} separates $G_E(Q^2)-f(Q^2)$ into plots for 
each spectrometer individually.  Here there is a gradual rise and
fall---albeit on the level of a tenth of a percent---in $G_E(Q^2)-f(Q^2)$ for Spectrometers A and C.  
For Spectrometer B, which is the workhorse at low $Q^2$, there is no such variation.  
Any deviations in the data from the continued fraction fit should show up in all three 
spectrometers if they are real.  Because this is not the case, the observed fluctuations likely are not 
intrinsic to $G_E$.  

We remind ourselves that the 4-parameter continued fraction fit to the full 1422-point Mainz data set
is good, and it is now somewhat improved by the renormalizations.  
After the renormalizations, $R_E$ does not change appreciably, but the $\chi^2$/dof decreases to about 1.4, which by some
measure is still too large. Consequently, we need to consider increasing  the size of the uncertainty limits.


\section{Final Results}
\label{sec:fr}

\begin{figure}[h]
\vskip -5 mm
\centerline{\includegraphics[width=9.2cm, angle=0]{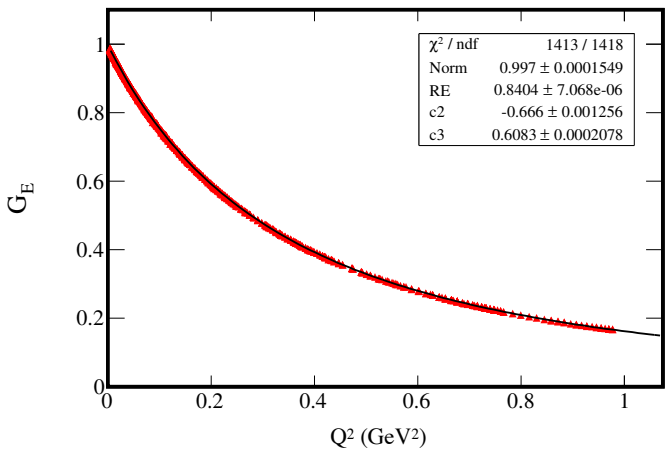}}
\vskip -9 mm
\caption{(Color online)
Final global fit with the 34 data sets renormalized and the point-to-point uncertainties on $G_E$ scaled up by 15\%.
}
\label{fig:FinalFit}
\end{figure}

The systematic deviations from the CF fit shown in Fig.~\ref{fig:DeltaABC} differ considerably from
spectrometer to spectrometer, suggesting that they are not intrinsic to $G_E(Q^2)$ and perhaps that
the point-to-point systematic uncertainties are underestimated.  Bernauer {\it et al.} themselves
have rescaled the uncertainties per normalization set by factors ranging from 1.07 to 2.3 \cite{Bernauer:2013tpr}.
Therefore, we repeated this exercise globally for $G_E(Q^2)$, and found that the uncertainties 
required rescaling by a factor 1.15.
Fig.~\ref{fig:FinalFit} shows the full data-set including our renormalizations of data 
sets and uncertainties.  
The resulting new fit (Fig.~\ref{fig:FinalFit}) has a $\chi^2$/dof\ of unity for the 
full data set.  All of the modifications we have made to the 
data have not changed $R_E$ more than a few parts per 
thousand. We obtain the value $R_E = 0.8404$ fm,  from this procedure, 
with a diagonal uncertainty of 0.00007 fm. 
The value  of $R_E$ fm remains  
consistent with the muonic hydrogen Lamb shift measurements.

\begin{figure}[h]
\vskip -3 mm
\includegraphics[width= 9.2 cm]{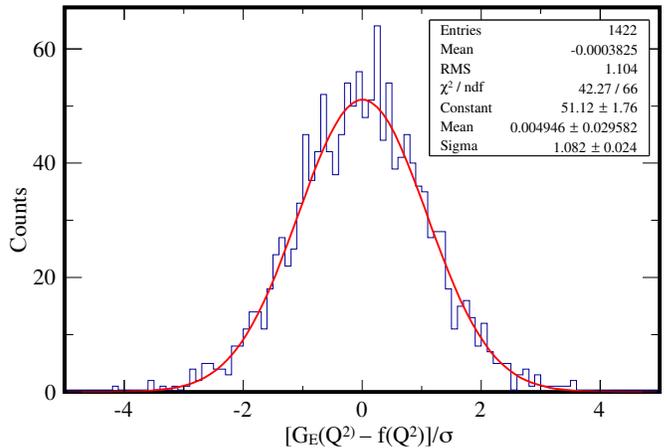}
\vskip -6 mm
\caption{(Color online)
Distribution of $[G_E(Q^2_i)-f(Q^2_i)]/\sigma_i$ for all data points.  Here the individual 
uncertainties on each point, $\sigma_i$, 
have all been rescaled by a factor of 1.15.
}
\label{fig:ErrorDist}
\end{figure}

Regarding the size and distribution of the uncertainties, the across-the-board increase of the 
uncertainty limits on $G_E$ by $15\%$ yields a normal distribution for $[G_E(Q^2_i)-f(Q^2_i)]/\sigma_i$.  
Fig.~\ref{fig:ErrorDist} shows the resulting histogram of this quantity for all 1422 points in the data set.  
A Gaussian fit
yields a mean of zero and a standard deviation of 1 with a good $\chi^2$/dof,
as expected for Gaussian statistics.

The statistical uncertainty from the $\chi^2$ fit is  small, and the overall uncertainty on $R_E$ 
is dominated by systematics.  We have estimated the systematic uncertainties by finding the spread
among a set of extracted radii: 1) Spectrometer B, $Q^2<0.02$ GeV$^2$, fits  to terms up to $Q^6$ 
constrained by assuming an exponential, Gaussian, or empirical charge distribution: 
0.836, 0.849, and 0.859 fm, respectively;  2)  $Q^2<1.0$ GeV$^2$, fits to the double dipole, 
continued fraction, and inverse polynomial fit forms: 0.830, 0.840, and 0.870 fm, respectively; and 
3)  a global fit to the 1422 points 
with each of the 34 normalization constants as free parameters, 0.827 fm (not reported in detail here).  
The average and standard deviation within this set
are $0.844$ fm and $0.016$ fm.  
We take this standard deviation as an estimate of the systematic uncertainty on $R_E$
from fit-model dependence, and keeping the central value from our previous analysis, conclude
that $R_E=0.840\pm 0.001_{\text stat}\pm 0.016_{\rm syst}$.


\vskip-2mm
\section{Conclusions}
\label{sec:conclusions}
\vskip-3mm


The Mainz data set is of extremely high quality---expansive, accurate, and self-consistent.  
We began with an analysis of the low-$Q^2$ part of the data set, where a polynomial expansion 
of the form factors should converge, and which should and does yield an accurate result 
for the proton radius.  We found a proton radius in 
agreement with the muonic hydrogen Lamb shift results and significantly smaller than the CODATA value.

We also analyzed the full data set,  assuming that $G_E$ is monotonically falling and inflectionless, 
and used a continued fraction form to map this.  
We rescaled the different data sets on a level that is smaller than the original normalization uncertainties.  
We  inflated the point-to-point
systematic uncertainties by $15\%$, which is well within reasonable systematic uncertainties for 
such an electron scattering experiment.  
We can then fit all data nicely using only 4 parameters.  This results in a $\chi^2$/dof\ of unity, 
and with some further consideration of other ways to fit the data, determine a proton radius 
$R_E = 0.840\pm 0.016$ fm. 

This result is in excellent agreement with the muonic Lamb shift results.  
Of course, if this is true, the proton is less interesting than we had hoped, and 
beyond-the-standard-model explanations of the proton radius puzzle will not be needed.   
One way to solve the current conundrum using scattering data
is to have independent confirmation of the shape 
of $G_E$ from other electron or muon scattering measurements, and a number are underway or 
in planning~\cite{Gilman:2013eiv,Gasparian:PRADJLAB,Mihovilovic2013,DistlerGriffioen}.  
In addition, a host of other relevant experiments are also underway or under analysis, 
including the completed but not yet published measurements of nuclear radii in other muonic 
atoms~\cite{CREMAnote}, and new high precision atomic level splitting experiments that will 
yield new and precise measurements of the proton 
radius~\cite{VuthaHessels2012,Beyer:2013JPhCS.467a2003B,Arnoult:2010,Flowers:2007}.  
We eagerly await all the new measurements that can elucidate the proton radius quandary.

\begin{acknowledgments}
We thank Jan Bernauer and Michael Distler for freely sharing their excellent published data, 
Douglas Higinbotham and Thomas Walcher for useful conversations, and 
Siyu Meng for performing fits using Mathematica.  
CEC thanks the National Science Foundation for support under grants PHY-1205905 and PHY-1516509 and KG and SM
thank the Department of Energy for support under grant DE-FG02-96ER41003.

\end{acknowledgments}

\bibliography{ProtonRadius}
\bibliographystyle{apsrev}

\end{document}